\documentclass[twocolumn,prl,preprintnumbers,superscriptaddress,amsmath,amssymb,english]{revtex4}
\usepackage{amsmath,amssymb,amsfonts,mathrsfs}
\usepackage{amsthm}
\usepackage{CJK}
\usepackage{bm}
\usepackage{babel}
\usepackage{graphicx}
\allowdisplaybreaks
\usepackage{braket}
\usepackage[breaklinks]{hyperref}
\usepackage{color}
\usepackage{epstopdf}
\hypersetup{colorlinks=true, linkcolor=blue, citecolor=blue, filecolor=blue, urlcolor=blue}

\begin{document}

\title{The One-dimensional Chiral Anomaly and its Disorder Response}

\author{Zheng Qin}
\affiliation{Institute for Structure and Function $\&$ Department of Physics, Chongqing University, Chongqing 400044, People's Republic of China}
\affiliation{Chongqing Key Laboratory for Strongly Coupled Physics, Chongqing 400044, People's Republic of China}

\author{Dong-Hui Xu}
\affiliation{Institute for Structure and Function $\&$ Department of Physics, Chongqing University, Chongqing 400044, People's Republic of China}
\affiliation{Chongqing Key Laboratory for Strongly Coupled Physics, Chongqing 400044, People's Republic of China}
\affiliation{Center of Quantum Materials and Devices, Chongqing University, Chongqing 400044, People's Republic of China}

\author{Zhen Ning}
\email[]{zhenning@cqu.edu.cn}
\affiliation{Institute for Structure and Function $\&$ Department of Physics, Chongqing University, Chongqing 400044, People's Republic of China}
\affiliation{Chongqing Key Laboratory for Strongly Coupled Physics, Chongqing 400044, People's Republic of China}

\author{Rui Wang}
\email[]{rcwang@cqu.edu.cn}
\affiliation{Institute for Structure and Function $\&$ Department of Physics, Chongqing University, Chongqing 400044, People's Republic of China}
\affiliation{Chongqing Key Laboratory for Strongly Coupled Physics, Chongqing 400044, People's Republic of China}
\affiliation{Center of Quantum Materials and Devices, Chongqing University, Chongqing 400044, People's Republic of China}

\date{\today}

\begin{abstract}
The condensed-matter realization of chiral anomaly has attracted tremendous interest in exploring unexpected phenomena of quantum field theory. 
Here, we show that one-dimensional (1D) chiral anomaly (i.e., 1D nonconservational chiral current under a background electromagnetic field) can be realized in a generalized Su-Schrieffer-Heeger model where a single gapless Dirac cone occurs. Based on the topological Thouless pump and anomalous dynamics of chiral displacement, we elucidate that such a system possesses the half-integer quantization of winding number. Moreover, we investigate the evolution of 1D chiral anomaly with respect to two typical types of disorder, i.e., on-site disorder and bond disorder. The results show that the on-site disorder tends to smear the gapless Dirac cone. However, we propose a strategy to stabilize the half-integer quantization, facilitating its experimental detection. Furthermore, we demonstrate that the bond disorder causes a unique crossover with disorder-enhanced topological charge pumping, driving the system into a topological Anderson insulator phase.
\end{abstract}
\maketitle

\emph{{\color{magenta}Introduction.}}---The exploration of Dirac matter provides a promising avenue to investigate various fundamental physics in relativistic quantum field theory ~\cite{PhysRevLett.57.2967, RevModPhys.81.109, PhysRevB.85.195320, RevModPhys.90.015001,halfWN}. Specifically, a single Dirac cone associates with quantum anomaly, playing an essential role in formulating exotic topological states in modern condensed matter physics~\cite{PhysRevLett.51.2077, PA,CA1,CA2,PhysRevLett.53.2449} which is an interesting manifestation of the topological phenomena, such as quantum anomalous Hall effects~\cite{Haldane, PhysRevLett.114.256601, PhysRevLett.106.166802,PhysRevLett.123.226602} and topological magnetoelectric effects of axion electrodynamics~\cite{PhysRevB.78.195424,axion}. Recently, the novel topological semimetallic phase, known as quantum anomalous semimetal characterized by the half-integer (or fractional) topological invariant, was introduced in \rm{Ref.~\cite{bhzQASM}} and further explored in several works~\cite{PhysRevB.106.045111,PhysRevB.105.L201106,PhysRevB.107.125153}.
So far, there have been many theoretical and experimental efforts that devote to observing the physics related to quantum anomaly~\cite{PA-exp1,PA-exp2,PhysRevLett.129.096601,DTSong,CA-exp1,CA-exp2,CA-exp3,SCI-WEYL,SCI-chiral2015,Sci-chiral2019}, but the direct evidence of hallmark half-integer topological invariant is rather scarce. The main reason is that the Dirac fermions always come in pairs in a realistic system according to the Nielsen-Ninomiya no-go theorem~\cite{Nielsen-Ninomiya,Jan-smit}, strongly hindering the direct observation of quantum anomalous physics in experiments.

Very recently, a significant experimental progress of two-dimensional (2D) quantum anomaly, termed as parity anomaly, was made in a semi-magnetic topological insulator (TI) heterostructure~\cite{PA-exp1,PA-exp2}.
Comparing with parity anomaly occurred in even spatial dimensions, the analogous phenomenon in odd spatial dimensions is the so-called chiral anomaly~\cite{CA-hd}, which have also been attracting extensive interests. Typical example of this is the non-conservative chiral current of Weyl fermions under an electromagnetic field, which is considered as the origin of the topological magnetoelectric effects of axion electrodynamics, and negative magnetoresistance of Weyl semimetals~\cite{PhysRevB.83.205101,RevModPhys.90.015001,DTSong,CA-exp1,CA-exp2,CA-exp3}. Up to now, exotic phenomena related to the chiral anomaly of Weyl quasiparticle in condensed matter systems are the subject of intense studies~\cite{CA-exp1,CA-exp2,CA-exp3,SCI-WEYL,SCI-chiral2015,Sci-chiral2019}, but these studies have been mainly limited to three spatial dimensions. The one-dimensional (1D) chiral anomaly and especially its analogous condensed matter realization are largely unexplored. Considering recent encouraging advancements of 1D topological states~\cite{TAI-exp,PhysRevB.103.235110, PhysRevX.8.031045, PhysRevB.102.205425, gSSH-1,gSSH-2,gSSH-3,gSSH-5,c1d-1,c1d-2,c1d-3}, the exploration of 1D chiral anomaly and its related novel physics is highly desirable.

In this work, we show that the chiral anomaly can be realized in a 1D topological nodal system. Without loss of generality, we choose the generalized Su-Schrieffer-Heeger (SSH) model with the long range hopping, a paradigmatic model used to investigate the 1D topological states~\cite{gSSH-1,gSSH-2,gSSH-3,gSSH-5}. With the proper setting of dimerization, which is accessible in experiments, the 1D single gapless Dirac cone characterized by a half-integer winding number is present~\cite{bhzQASM}. 
This massless Dirac fermion gives rise to the nonconservation of chiral current under a background electromagnetic field, served as a signature of 1D chiral anomaly.
To facilitate the experimental measurement of half-integer quantization, we investigate the topological Thouless pumping and anomalous dynamics of chiral displacement~\cite{Thouless1983,NiuQ1990,Tpump-1,Tpump-2,MCD}. Moreover, it is worth noting that the influence of disorder in one spatial dimension is more remarkable than that in higher spatial dimensions  where the usual perturbation theory breaks down~\cite{PatrickLee,Mirlin-RMP}, causing another difficulty for the experimental detection of 1D chiral anomaly. Therefore, we investigate the evolution of 1D chiral anomaly with respect to disorder, including the on-site disorder and bond disorder. Remarkably, we find that the on-site disorder can be effectively suppressed and the half-integer quantization can be nearly preserved in the regime of moderate disorder strength. On the other hand, the interplay between disorder and topology can lead to unexpected phenomena. 
Thereby, we further investigate the case of bond disorder. The results show a unique crossover with disorder-enhanced topological charge pumping, and then the topological gapless state evolves into an exotic topological Anderson insulator (TAI) phase~\cite{TAIshen} with integer-quantized topological charge.

\begin{figure}[htb]
	\centering
	\includegraphics[width=0.47\textwidth]{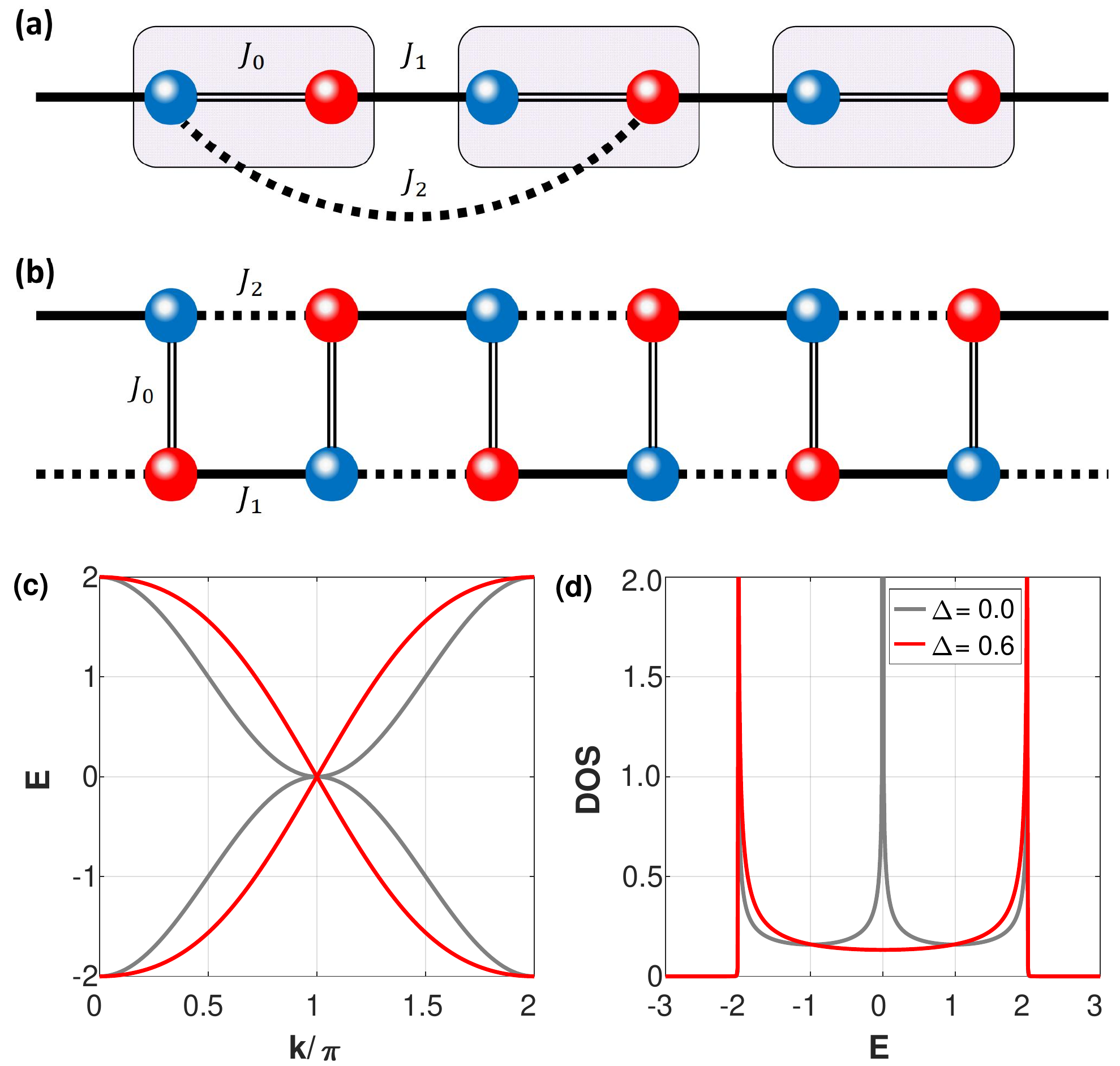}%
	\caption{ (a) The schematic diagram of a generalized SSH model with the nearest and next-nearest neighbor hopping, containing two sublattices A and B. (b) The equivalent structure in panel (a) with a two-leg ladder structure. Two SSH chains are dimerized with staggered hopping amplitudes $J_1$ and $J_2$, respectively. The coupling between two chains is described by $J_0$. (c) The band structures and (d) the corresponding density of states 
 with parameter
 $\Delta = 0$ and $\Delta = 0.6$.  }
\label{Fig:Fig1}
\end{figure}

\emph{{\color{magenta}The 1D chiral anomaly in generalized SSH model.}}--- We start with a generalized SSH model, as sketched in Fig.~\ref{Fig:Fig1}(a).  This model can also be viewed as a two coupled SSH chains system with a two-leg ladder structure [see Fig.~\ref{Fig:Fig1}(b)]~\cite{gSSH-1,gSSH-2,ladder-1,ladder-2,ladder-3}. Besides, we generally consider two typical types of disorder, i.e., the on-site disorder and bond disorder. The tight-binding Hamiltonian of the lattice model is given by
\begin{equation}
\begin{aligned}
H&=H_0+V,\\
H_0&=\sum_{i}(\psi^{\dag}_{i} T_{0}\psi_{i}+\psi^{\dag}_{i+1} T_1 \psi_{i}+h.c.),\\
V &=\sum_i(W^{a}_i a^{\dag}_{i}a_{i} + W^{b}_i b^{\dag}_{i}b_{i}) + \sum_i(U_i a^{\dag}_{i}b_{i} + h.c.),\\
\end{aligned}
\label{eq:H}
\end{equation}
with
\begin{equation}
\begin{aligned}
T_{0}&= \begin{pmatrix}
0 & J_0 \\
J_0 & 0 \\
\end{pmatrix},
T_1= \begin{pmatrix}
0 & J_1 \\
J_2 & 0 \\
\end{pmatrix},
\end{aligned}
\end{equation}
where the two component spinors $\psi^{\dag}_{i} = [a^{\dag}_{i},b^{\dag}_{i}]$  and $\psi_{i}$ are composed of creation and annihilation operators acting on the sublattice $(A,B)$ at unit cell $i$. The schematic illustration of the hopping matrices $T_{0,1}$ are shown in Fig.~\ref{Fig:Fig1}(b), where the hopping amplitudes between unit cell are $J_{1,2}$ and we set interchain hopping $J_0=1$ without loss of generality. The lattice constant and Planck constant are set to be unity $a=\hbar=1$ throughout the following discussion.
The on-site disorder is modeled by the two independent random numbers $W^{A}_i$ and $W^{A}_i$, whose values are taken from the interval $[-W/2,W/2]$. The parameter $U_i$ represents the bond disorder which takes value randomly from the interval $[-U/2,U/2]$. Here, $W$ and $U$ denote the strength of disorder in the unit of $J_0$.

\begin{figure}[htb]
	\centering
	\includegraphics[width=0.47\textwidth]{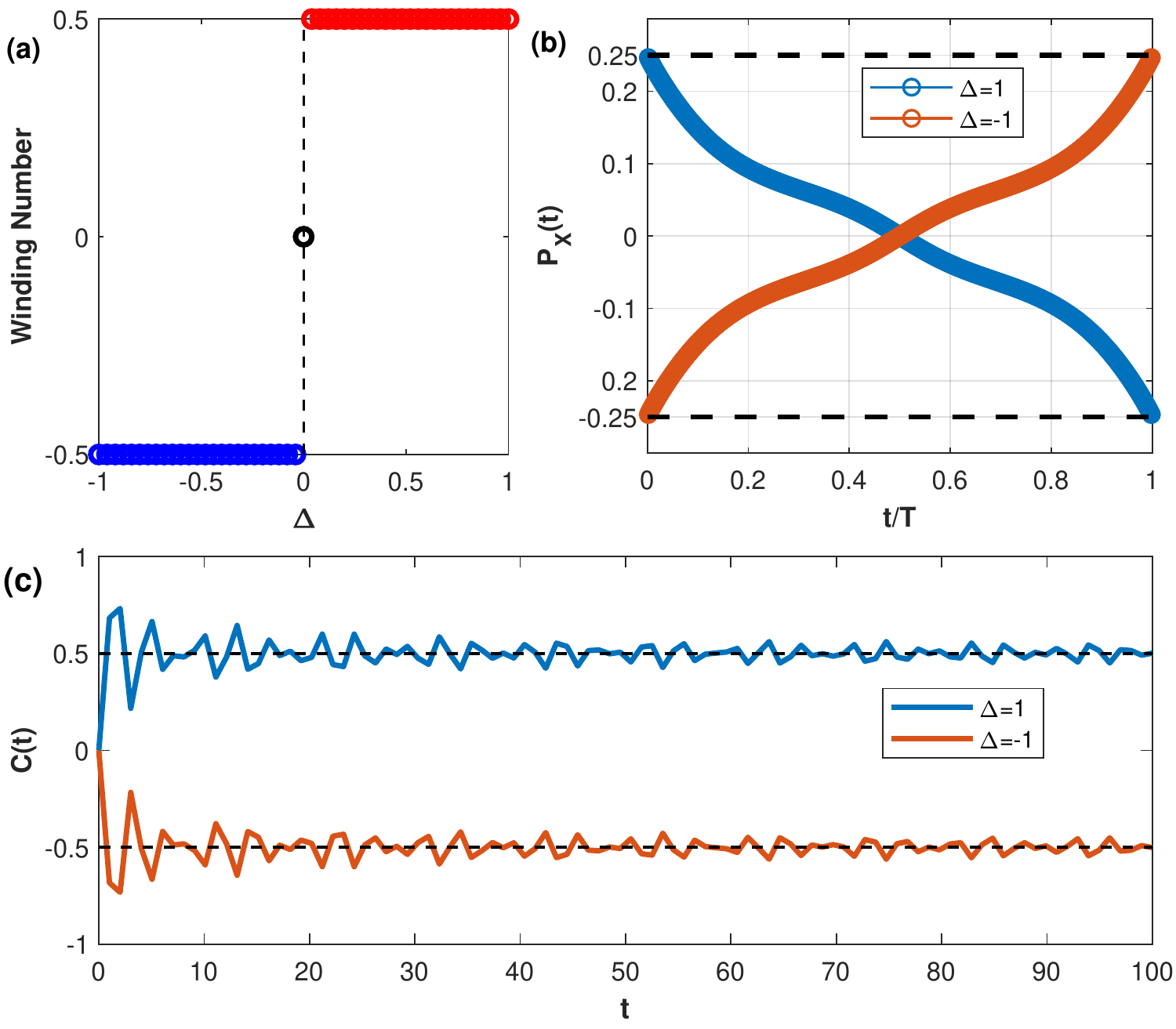}%
	\caption{ (a) The winding number of the gapless system Eq.~(\ref{eq:hk}) as a function of $\Delta$. (b) The time-evolution of the polarization $P_X(t)$ in one period $T$. The pumped charge can be read off from $Q = P(T) - P(0)$, and we have $Q= -\frac{1}{2}$ ($\Delta>0$) and $Q =\frac{1}{2}$ ($\Delta <0$). (c) The time-evolution of chiral displacement $C(t)$ which oscillate around the winding number $w=\pm \frac{1}{2}$.}
\label{Fig:Fig2}
\end{figure}

First of all, we consider the ideal system and set the disorder potential to zero, i.e., $W^{A}_i = W^{A}_i=U_i=0$. Without disorder, the lattice Hamiltonian Eq.~(\ref{eq:H}) can be transformed into the momentum space, and we have $H_0 = \sum_{k}\psi^{\dag}_{k}h_k\psi_{k}$, where $h_k$ can be expressed as
\begin{equation}
\begin{aligned}
h_k&= h_x(k)\sigma_x + h_y(k)\sigma_y
\end{aligned}
\label{eq:hk}
\end{equation}
with
\begin{equation}
\begin{aligned}
h_x(k) &= J_0 + (J_1+J_2)\cos(ka),\\
h_y(k) &= (J_1-J_2)\sin(ka),
\end{aligned},
\end{equation}
and $\sigma_{x,y}$ are Pauli matrices and act on the sublattice $(A, B)$. For convenience, we redefine $2J = J_1+J_2$ and $2\Delta = J_1-J_2$, and thus the parameter $\Delta$ characterizes the lattice dimerization of a SSH chain. 
Here, we focus on the single gapless node in the Brillouin zone, which can be realized by fixing $J = \frac{1}{2}$. We plot the band structure and the corresponding density of states (DOS) in Figs.~\ref{Fig:Fig1}(c) and \ref{Fig:Fig1}(d), respectively. The dimerization parameter $\Delta$ strongly affects the feature of band dispersion. When $\Delta = 0$, we have a quadratically gapless nodal point at $k = \pi$ and its DOS is divergent at the band center. Such a singularity of DOS can be removed by introducing the dimerization (i.e., $\Delta \neq 0$) and the DOS becomes a finite value.  Significantly, the presence of lattice dimerization gives rise to a single gapless Dirac cone with linear dispersion [see Fig.~\ref{Fig:Fig1}(c)], which may invoke a topological transition.
To reveal the process, we employ Eq.~(\ref{eq:hk}) to calculate the winding number as~\cite{ladder-3}   
\begin{equation}
\begin{aligned}
w = \int^{\pi}_{-\pi}\frac{dk}{2\pi}\frac{h_x\partial_kh_y-h_y\partial_kh_x}{h^2_x+h^2_y}.
\end{aligned}
\label{eq:wn}
\end{equation}
After integrating using Eq.~(\ref{eq:wn}), we obtain the winding number $w = \frac{1}{2}\mathrm{sgn}(\Delta)$~\cite{bhzQASM}, indicating that the 1D gapless system with a single Dirac cone carries half-integer winding number as long as any finite dimerization occurs, i.e., $\Delta \neq 0$ [see Fig.~\ref{Fig:Fig2}(a)]. It is worth noting that there is no singularity in the integrand at the nodal point~\cite{SM}). To reveal that the massless Dirac fermion leads to the nonconservation of chiral current under a background electromagnetic field, we expand the Hamiltonian Eq.~(\ref{eq:H}) near the nodal point ($k_0= \pi$) and obtain the continuous Dirac Hamiltonian with linear dispersion as $H_{Dirac}= \sum_{q}\psi^{\dag}_{q}v_F q\sigma_{y}\psi_{q}$, where $v_F = 2\Delta$ defines the velocity and $q$ represents the wave vector from the nodal point. The corresponding relativistic Lagrangian in (1+1) space-time can be written as $L=\overline{\psi}\gamma^{\mu}\partial_{\mu}\psi$ (see details in the Supplementary Material~\cite{SM}). According to Fujikawa's derivation of chiral anomaly ~\cite{fujikawaBOOK,Fujikawa}, we consider a background gauge field where $\partial_{\mu} \rightarrow D_{\mu} = \partial_{\mu} + eA_{\mu}$. In this case, the conservation of chiral current $j^5_{\mu} = \overline{\psi}\gamma^{\mu}\gamma^5\psi$ breaks down at the quantum level $\partial_{\mu}j^5_{\mu} = \frac{e}{2\pi}\epsilon_{\mu\nu}F^{\mu\nu}$~\cite{fradkinQFT}. Therefore, the generalized SSH model with a single massless Dirac fermion carrying half-integer winding number provides a realistic platform to realize the 1D chiral anomaly, similar to the half-quantized Chern number (Hall conductance) of parity anomaly in two spatial dimensions~\cite{Haldane}.

The 1D chiral anomaly with half-integer winding number can be revealed by the topological Thouless charge pump in experiments~\cite{Tpump-1,Tpump-2}. For a periodically driven system, the transported charge during one period $T$ can be evaluated by the time evolution of polarization as~\cite{Resta1988,Smith1993}
\begin{equation}
\begin{aligned}
Q = P_{\mathrm{\hat{X}}}(T)-P_{\mathrm{\hat{X}}}(0)= \mathrm{\frac{1}{2\pi}Im\log[\langle\Psi(t)|e^{i2\pi \hat{X}/L}|\Psi(t)\rangle]},
\end{aligned}
\label{eq:tpQ}
\end{equation}
where the instantaneous wave function $|\Psi(t)\rangle$ is composed of occupied states at time $t$ and $\hat{X}$ is the position operator. For a lattice with $L$ unit cells, the operator  $\hat{X}$ can be written in the form of diagonal matrix as $\hat{X} = \mathrm{diag}\{-L,-L+1,\cdots,L\}\otimes \sigma_0$. To investigate the charge pumping, we modulate the static Hamiltonian Eq.~(\ref{eq:H}) by a time periodic potential~\cite{bhzQASM} $H'(t) = J_0\sum_{i}[\psi^{\dag}_{i} \sin^2(\pi t/T)\sigma_x\psi_{i} +\psi^{\dag}_{i} \sin(2\pi t/T)\sigma_z\psi_{i}]$. As shown in Fig.~\ref{Fig:Fig2}(b), the transfer of pumped charge in one period exhibits one half of the elementary charge, i.e., $|Q| = \frac{1}{2}$, in contrast to the gapped topological phase where the pumped charge is always an integer.
%
Such topological property with half-integer quantization can also be detected in experiments by measuring the mean chiral displacement (MCD)~\cite{MCD,TAI-exp}. For the lattice Hamiltonian Eq.~(\ref{eq:H}), we can calculate dynamics of the chiral displacement operator
\begin{equation}
\begin{aligned}
\mathrm{C}(t) = \langle\varphi(t)|(2\hat{\Gamma} \hat{X})|\varphi(t)\rangle
\end{aligned},
\label{eq:Ct}
\end{equation}
where the $\hat{\Gamma}$ is the chiral operator defined as $\Gamma = \mathrm{diag}\{1,\cdots,1\}\otimes \sigma_z$. Choosing an arbitrary initial wave function $|\varphi_0\rangle$, its time evolution is given by $|\varphi(t)\rangle = \exp(-iH_0t)|\varphi_0\rangle$. The dynamics $\mathrm{C}(t)$ displays a oscillatory behavior and converges to the winding number $w$ as shown in Fig.~\ref{Fig:Fig2}(c). 
By taking the time average, the MCD is given by $\mathrm{\bar{C}} = \lim_{t\rightarrow\infty}\frac{1}{t}\int^{t}_0 d t' \mathrm{C}(t')=\frac{1}{2}$, confirming the presence of 1D chiral anomaly in the generalized SSH model with single massless Dirac fermions. Otherwise, the long time dynamics of $\mathrm{C}(t)$ is trivial in the absence of dimerization with $\Delta=0$.
Moreover, due to the inevitable existence of disorder in realistic systems, the evolution of 1D chiral anomaly with respect to disorder requests to be further verified.

\begin{figure}[htb]
	\centering
	\includegraphics[width=0.47\textwidth]{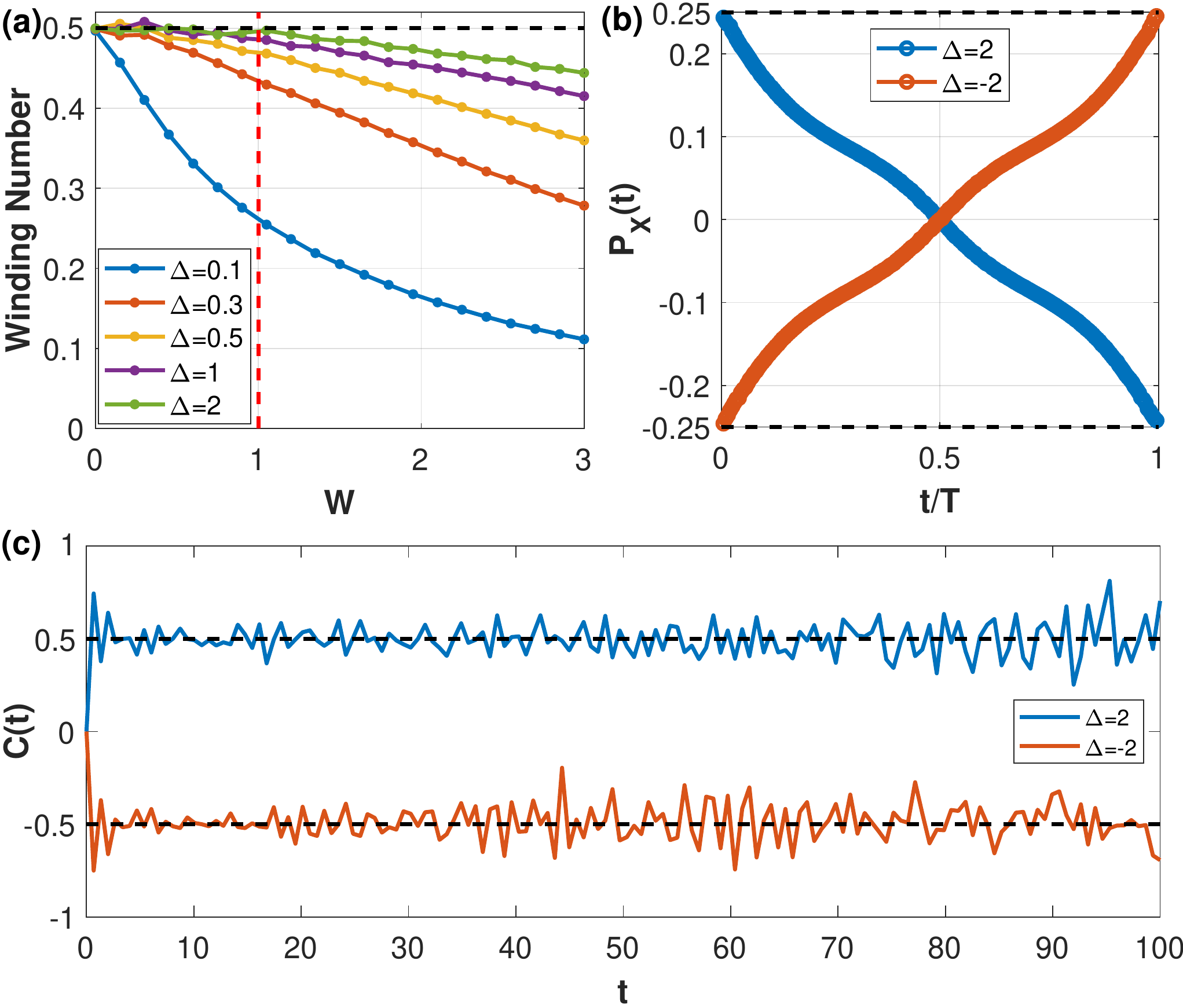}%
	\caption{ (a) The disorder-averaged winding number as a function of the strength of on-site disorder $W$ for different values of $\Delta$. For $W = 1$ (denoted by the vertical dashed line), the winding number can be can nearly be preserved as $\frac{1}{2}$ with increasing $\Delta$. (b) The disorder-averaged time-evolution of polarization $P_X(t)$ in one period $T$. (c) The time-evolution of chiral displacement $C(t)$ for the system with $L=2000$ unit cells under on-site disorder. The strength of disorder is fixed to $W =1$ in panels (b) and (c).}
\label{Fig:Fig3}
\end{figure}

\emph{{\color{magenta}Stability under on-site random fluctuations.}}---We then consider the influence of random fluctuation with on-site disorder described in Eq.~(\ref{eq:H}).
The random potential breaks the translational symmetry of lattice, 
so the winding number cannot be calculated in momentum space. Here, we alternatively construct the projection operators based on diagonalizing the real space tight-binding Hamiltonian Eq.~(\ref{eq:H}),
and then the winding number in real space can be given by~\cite{realWN,TAI-exp}
\begin{equation}
\begin{aligned}
w = -\hat{Q}_{BA}[\hat{X},\hat{Q}_{AB}]
\end{aligned},
\label{eq:wnr}
\end{equation}
where $\hat{Q}_{AB} = \hat{S}_A \hat{Q} \hat{S}_B$, $\hat{Q} = \hat{S}_{+} - \hat{S}_{-}$ with the positive and negative energy projection operators $\hat{S}_{\pm}$, and $\hat{S}_{A,B} = \frac{1}{2}(\hat{I}\pm\hat{\Gamma})$ denote the sublattice projection operators. To reveal the influence of on-site disorder on the 1D generalized SSH system with single Dirac cone, we compute the disorder-dependence of winding number using Eq.~(\ref{eq:wnr}) for different values of $\Delta$ as depicted in Fig.~\ref{Fig:Fig3}(a). It is found that the half-integer winding number is destroyed since the chiral symmetry is broken in the presence of on-side disorder. Fortunately, the enhancement of dimerization is in favour of stabilizing the half-quantized feature against disorder as shown Fig. S2 in the Supplemental Material ~\cite{SM}. For instance, the winding number can nearly be preserved as $\frac{1}{2}$ up to a relatively strong disorder strength $W\approx1$ when $\Delta = 2$ [see the blue-doted line in Fig.~\ref{Fig:Fig3}(a)]. This can also be verified by the one half of pumped charge in the topological Thouless pumping [see Fig.~\ref{Fig:Fig3}(b)]. As shown in Fig.~\ref{Fig:Fig3}(c), we depict the time evolution of chiral displacement $\mathrm{C}(t)$. Although the fluctuation of $\mathrm{C}(t)$ is larger that in the absence of disorder, the time average of $|\mathrm{C}(t)|$ can always converge to $w\approx \frac{1}{2}$ in the regime of moderate disorder
strength if increasing the dimerization of system (see Fig. S1 in the Supplemental Material~\cite{SM}). Therefore, despite the fact that half-quantization of winding number is not exact, 
the interplay between the disorder and dimerization offers a strategy for effectively suppressing the influence of on-site disorder, which facilitate the observation of 1D chiral anomaly in experiments.


\begin{figure}[htb]
	\centering
	\includegraphics[width=0.47\textwidth]{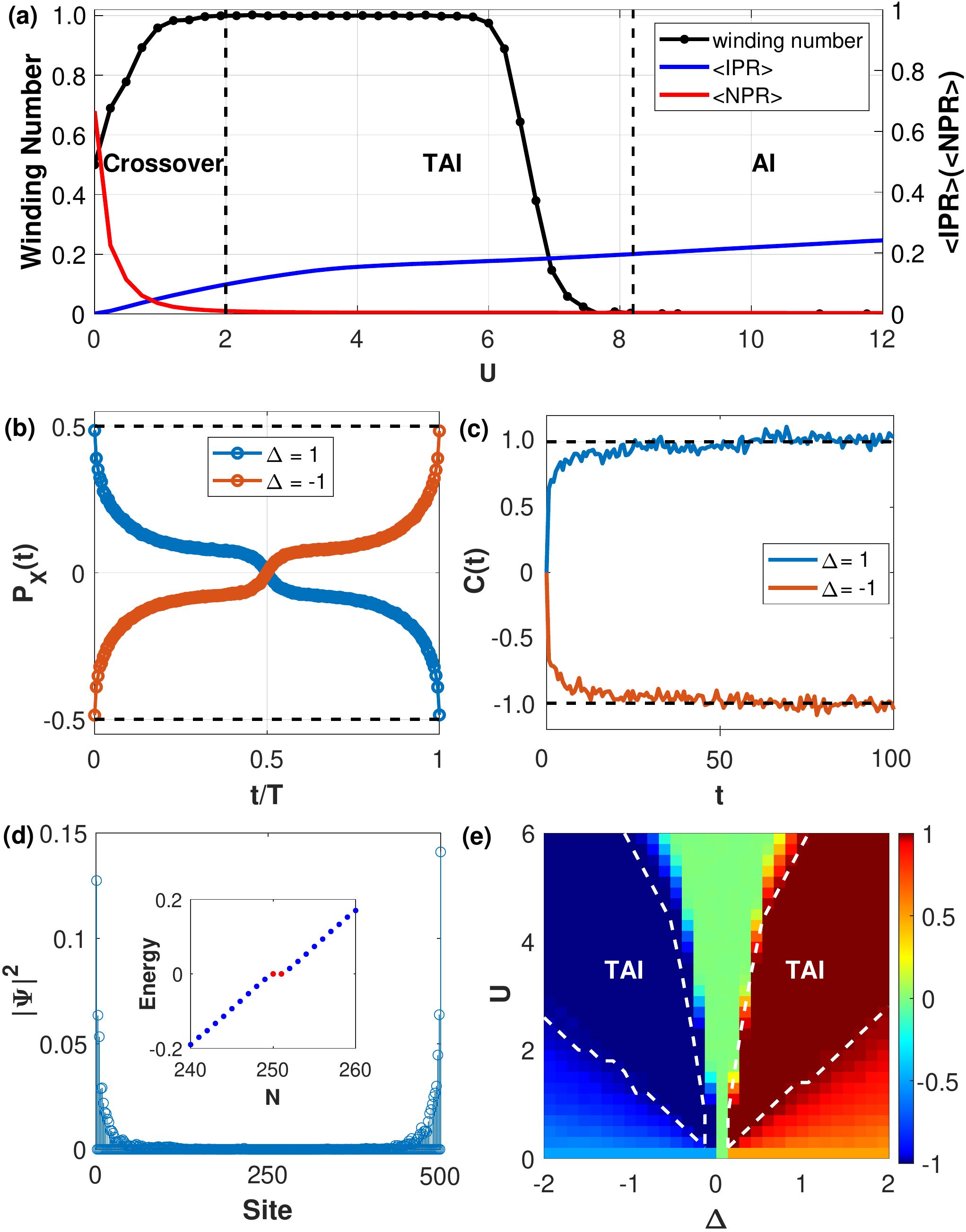}%
	\caption{ (a) The disorder-averaged winding number (left axis) and the averaged IPR/NPR (right axis) as a function of strength of random bond disorder $U$. Three topologically different regimes are labeled by crossover ($w<1$, $\rm{IPR}\ne0$, and $\rm{NPR}\ne0$), the TAI ($w=1$, $\rm{IPR}\ne0$, and $\rm{NPR}=0$) and the trivial Anderson Insulator ($w=0$, $\rm{IPR}\ne0$, and $\rm{NPR}=0$). (b) The disorder-averaged time-evolution of polarization $P_X(t)$. The pumped charge in one period $T$ is integer quantized $|Q| = 1$. (c) The time-evolution of chiral displacement $C(t)$. (d) The probability density of the boundary wave function in the TAI phase. The inset show the energy spectrum near the band center $E=0$, and two zero modes are denoted by the red dots. (e) The phase diagram depicted by $\Delta$ and $U$, where the regimes of TAI are enveloped by the white dashed line. We set the parameter $\Delta=1$in panels (a)-(d) and the disorder strength $U=4$ in panels (b)-(d).}
\label{Fig:Fig4}
\end{figure}

 \emph{{\color{magenta}Topological transition of 1D chiral anomaly induced by bond disorder.}}---As discussed above, we can see that the on-site disorder cannot induce any topological transition. To establish the connection between the 1D chiral anomaly and its accompanied topological transition, we further investigate the effects of bond disorder. As shown in Fig.~\ref{Fig:Fig4}(a), with the increase of disorder strength, the calculated wind number displays from the initial $w=\frac{1}{2}$ to an another topological phase with $w=1$ undergoing a crossover with the disorder-enhanced topological charge pumping, and finally evolves to a trivial phase with $w=0$. To reveal the interplay between the topological transition and disorder-induced localization of states, we calculate the inverse participation ratio (IPR) and the normalized participation ratio (NPR)~\cite{IPR/NPR,Mirlin-RMP}, which can be obtained by
 \begin{equation}
\begin{aligned}
\mathrm{IPR} = \sum_{i,n}|\varphi^i_n|^4, \quad \mathrm{NPR} = (L\sum_{i,n}|\varphi^i|^4)^{-1},
\end{aligned}
\label{eq:INPR}
\end{equation}
where $\varphi^i_n = \langle i |\varphi_n\rangle$ is the wave function at site $|i\rangle$ for the $n$-th eigenvector $|\varphi_n\rangle$. The calculated results of IPR and NPR are plotted as the blue and red lines in Fig.~\ref{Fig:Fig4}(a). In the absence of disorder, the system is in the extended phase with $\mathrm{IPR}=0$ and $\mathrm{NPR}\ne 0$. In the case of low disorder strength, we can see a crossover which corresponds to an intermediate regime where the IPR and NPR are both finite. When the disorder strength increases, the NPR decays rapidly and there is only the finite IPR. As a result, the wave functions are localized and the system evolves into the Anderson insulator phase with $w=1$~\cite{SM}. This topological regime corresponds to a TAI phase~\cite{TAIshen} which belongs to the chiral $\mathrm{AIII}$ symmetry class~\cite{realWN,halfWN}. To further confirm this, we respectively calculate the topological Thouless charge pump and dynamics of the chiral displacement as shown in Figs.~\ref{Fig:Fig4}(b) and ~\ref{Fig:Fig4}(c). Besides, the TAI phase should host the related boundary states, and thus we calculate the energy spectrum using open boundaries with 500 unit cells. As plotted in Fig.~\ref{Fig:Fig4}(d), we can see that there are two zero-energy modes whose wave functions are localized at the boundary sites. Moreover, it is worth noting that, similar to the case of on-site disorder, the dimerization parameter $\Delta$ may affect the critical disorder strength of topological transition. To reveal this, we calculate a comprehensive phase diagram depicted by the parameters $(\Delta,U)$ as plotted in Fig.~\ref{Fig:Fig4}(e).  It is found that the TAI cannot be formed by bond disorder if the dimerization parameter is absent $\Delta = 0$. Besides, we can see that the region of crossover is expanded with increasing $\Delta$.
\emph{{\color{magenta}Summary.}}--- In summary, we have theoretically investigated the 1D chiral anomaly in the generalized SSH model with long range hopping. We argue that such topological system characterized by the winding number $w=\frac{1}{2}$  possesses a single gapless Dirac cone, which can give rise to the nonconservation of chiral current under a background electromagnetic field. We show that the 1D chiral anomaly leads to the half-quantized topological charge Thouless pumping and anomalous dynamics of chiral displacement. Through controlling the dimerization of SSH chains, we uncover that the influence of on-site random fluctuation upon the experimental detection can be effectively minimized and the half-integer quantization can be nearly preserved in the regime of moderate disorder strength. 
Moreover, under the bond disorder, we show that the system with 1D chiral anomaly can evolve into a TAI phase with integer-quantized topological charge, which is resulting from a unique disorder-enhanced topological Thouless pumping.
Considering the recently experimental progress in studies of various 1D topological phases in artificial and condensed matter systems~\cite{opl,realz1,realz2,realz-haldane,TAI-exp,AL,FTAI}, we expect that our proposed 1D chiral anomaly in generalized SSH model can be realized in experiments and further attract more intense studies of novel lower-dimensional topological physics.

\emph{{\color{magenta}Acknowledgments.}}--- The authors thank Dr. Bo Fu for useful discussions. This work was supported by the National Natural Science Foundation of China (NSFC, Grants No. 11974062, No. 12222402, No. 12147102, and No. 12074108), and the Beijing National Laboratory for Condensed Matter Physics.

\bibliographystyle{apsrev4-2}
\bibliography{CASM-ref}

\end{document}